\documentclass[pre,twocolumn,superscriptaddress,showpacs]{revtex4}

\usepackage[dvips]{graphicx}
\usepackage{amsmath}

\begin{document}

\title{Persistence in systems with conserved order parameter}

\today

\author{P. Gonos and A. J. Bray \\ \medskip {\em Department of Physics 
and Astronomy, University of Manchester, Manchester M13 9PL, U.K.}}

 
\begin{abstract}

We   consider   the   low-temperature   coarsening   dynamics   of   a
one-dimensional   Ising  ferromagnet   with   conserved  Kawasaki-like
dynamics   in  the  domain   representation.   Domains   diffuse  with
size-dependent  diffusion  constant,   $D(l)  \propto  l^\gamma$  with
$\gamma =  -1$. We  generalize this model  to arbitrary  $\gamma$, and
derive  an expression for  the domain  density, $N(t)  \sim t^{-\phi}$
with   $\phi=1/(2-\gamma)$,  using  a   scaling  argument.    We  also
investigate    numerically   the    persistence    exponent   $\theta$
characterizing  the  power-law  decay  of  the  number,  $N_p(t)$,  of
persistent  (unflipped)  spins at  time  $t$,  and find  $N_{p}(t)\sim
t^{-\theta}$  where $\theta$  depends  on $\gamma$.  We  show how  the
results for $\phi$ and $\theta$ are related to similar calculations in
diffusion-limited  cluster-cluster aggregation  (DLCA)  where clusters
with  size-dependent diffusion  constant diffuse  through  an immobile
`empty' phase  and aggregate  irreversibly on impact. Simulations show 
that, while $\phi$ is the same in both models, $\theta$ is different 
except for $\gamma=0$.  We  also  investigate models that interpolate 
between symmetric domain diffusion and DLCA.

\end{abstract}



\maketitle

\section{Introduction}

Phase  separation  in  the  one-dimensional  Ising  ferromagnet  is  a
well-understood phenomenon \cite{bray94,derrida94,derrida95}.  At zero
temperature, with  Glauber (nonconserved) dynamics,  coarsening occurs
through the  irreversible annihilation of  domain walls, and  the mean
domain size  grows as $L(t)  \sim t^{1/2}$. The coarsening  process is
equivalent   to  the  diffusion-limited   single-species  annihilation
reaction  $A+A\rightarrow  0$, where  $A$  particles represent  domain
walls.   Although no  simple  particle model  can  be constructed  for
Kawasaki (spin-exchange) dynamics due to the nearest-wall correlations
introduced  by  the   conservation  principle,  qualitatively  similar
coarsening dynamics kinetics (but  with a different domain-growth law,
$L(t) \sim t^{1/3}$) have been observed for spin-exchange.

Results  have been  obtained  for the  case  of non-conserved  Glauber
dynamics for the  growth law governing the asymptotic  increase of the
average domain length $L(t)$ and recently even an exact expression for
the  `persistence' exponent  $\theta$ \cite{derrida95}.  This exponent
regulates  the  asymptotic  decay,  $Q(t) \sim  t^{-\theta}$,  of  the
persistence  probability, which  is the  probability that  a  spin has
never flipped (or, equivalently, the  fraction of spins which have not
flipped) up to  time $t$. It is a new  dynamical exponent, not related
to the domain-growth exponent \cite{majumdar}.

For conserved Kawasaki dynamics much  less is known. It has been shown
that  the  average domain  length  grows  algebraically  with time  as
$L(t)\sim  t^{1/3}$ instead of  $\sim t^{1/2}$,  a consequence  of the
$1/l$ dependence that emerges  at low temperature, from the underlying
spin-exchange  processes, for  the effective  diffusion constant  of a
domain  of   length  $l$.  Cornell  and   Bray  \cite{cornell96}  have
demonstrated that this size-dependence  vanishes in the presence of an
external driving  force. A calculation of the  domain density, $N(t)$,
gave  $N(t)\sim t^{-1/2}$ as  expected for  size-independent dynamics.
We will present a similar  calculation for the unbiased Kawasaki model
to obtain  an expression  for the density  which we compare  with data
from numerical simulations.

To  the  best  of  our   knowledge  no  analytical  treatment  of  the
persistence  probability exists  in  the literature  for the  Kawasaki
problem.  The  only numerical simulation  of the system appears  to be
\cite{sire98}  where  the  value   of  the  exponent  is  measured  as
$\theta=0.73$.  We  have repeated  this study for  a larger  sample of
initial configurations to obtain a similar value of $\theta$, and have
also simulated  the same system with the  general domain-diffusion law
$D(l)  \sim l^{\gamma}$  for a  range of  negative,  integer $\gamma$.
These models are not realisations of spin-exchange but are nonetheless
perfectly  legitimate in  the domain  description and  are furthermore
closely  related to  irreversible cluster-cluster  aggregation (DLCA),
another conserved system.  We will show that our  numerical values for
the dynamic  exponent $\phi$ in this generalised  Kawasaki model match
the expression  obtained by Helen  et al.\ \cite{hellen01}  for DLCA,
i.e.  \  $\phi^{cc}=1/(2-\gamma)$,  in  the range  that  we  examined.
Interestingly,  the  broken  symmetry in  cluster-cluster  aggregation
between the `filled'  and `empty' phases appears to  have no effect on
the  rate of  domain growth.   It  does, on  the other  hand, lead  to
differences between  $\theta^{cc}=2/(2-\gamma)$ (where the superscript
$cc$ indicates cluster-cluster aggregation) and our own numerical data
which yield, for $\gamma\le  -2$, lower limits on $\theta(\gamma)$ in
the symmetric spin system.

The  structure of  the paper  is as  follows. In  the next  section we
examine  Kawasaki spin-exchange  by  reducing the  dynamics to  domain
diffusion in a  suitable time-frame, and use a  scaling hypothesis for
the size distribution  of domains at time $t$,  $P(l,t)$, to calculate
an  expression for  the domain  density  in this  regime.  In  section
\ref{persprob} we  present data from  simulations of this  model which
indicate  that the persistence  exponent $\theta\simeq  0.71$. Section
\ref{aggg} introduces the other conserved system that will concern us,
irreversible cluster-cluster  aggregation, and contains  a comparative
treatment  of both  systems as  well as  systems  with characteristics
intermediate  between  these  two  models.  We  begin  by  introducing
cluster-cluster  aggregation and  a few  useful known  results through
which  the  correspondence   between  Kawasaki  domain  diffusion  and
aggregation cluster  diffusion will be assessed, and  continue with an
examination of  the various exponents for  weak (Sec.\ref{gamma1}) and 
strong  (Sec.\ref{gamma2})  size-dependence of the cluster (or domain) 
diffusion constant.

\section{The Kawasaki-Ising model}
\label{kawa}

Unfortunately,   a    straightforward   reduction   to    a   particle
reaction-diffusion system,  such as is  the case for  Glauber dynamics
and  the  single-species annihilation  process,  is  not possible  for
conserved  dynamics.   Exact  mapping  is retrieved  in  the  infinite
temperature limit, where  the walls behave like branching-annihilating
fixed particles \cite{avraham}, but is rather complicated to follow at
any finite  $T$. In  the low-temperature limit  of interest  here, the
simplest  description  of Kawasaki  dynamics  is  in  terms of  domain
diffusion  \cite{cornell91,krapivsky98}   which  we  describe  briefly
below.

The  three  spin  processes  that   we  wish  to  encapsulate  in  our
higher-level description are:
\begin{enumerate}
\item $-+-- \rightarrow --+-$ (spin diffusion)
\item $-+-+ \rightarrow --++$ (spin capture)
\item $--++ \rightarrow -+-+$ (spin emission)
\end{enumerate}
where it  is clear that  capture is just  the reverse of  emission and
hence  $\Delta  E_{3}=-\Delta  E_{2}$  (and  $\Delta  E_{1}=0$).   The
corresponding rates differ by  factors of order $\exp{(-4J/T)}$ so, in
the limit  $T \ll J$, the  time-scales for diffusion  and emission are
widely separated, with capture occurring instantaneously.

Implementing  spin-exchange dynamics  in this  way on  a ferromagnetic
lattice from  a random initial  configuration leads initially  to spin
diffusion,  followed by the  capture of  all available  isolated spins
until the latter disappear from the system. The rate of capture decays
exponentially  with time  and contributes  little  towards coarsening,
whatever the  size-distribution of domains at  $t=0$.  Thereafter, the
system  becomes  trapped  in  a  metastable state  whose  lifetime  is
extremely  long   and  diverges   to  infinity  at   zero  temperature
\cite{godreche02}.  At  any finite temperature,  the metastability can
only be  broken once a  rare splitting event  occurs to provide  a new
isolated  spin and  even then  the system  refreezes quickly  once the
issued spin is absorbed.

Spins produced  in this way  behave like random walkers,  initially at
$x=1$, surrounded by two absorbing  boundaries at $x=0$ and $x=l$, for
a  spin emitted  into domain  of length  $l$. The  spin  is eventually
captured by  either of the  boundaries with probabilities  $1-1/l$ and
$1/l$ respectively. Absorption at  $x=l$ results in the domain, across
which  the  spin  has  diffused,  being  moved  by  one  step  whereas
absorption at  the other boundary  returns the system to  its original
configuration. The combined  rate for one step of  the domain is $\sim
l^{-1}\exp{(-4J/T)}$ and it is this  overall motion of the domain that
facilitates coarsening close to $T=0$ \cite{cornell91}.

If $t_p$ is  the physical time, in order to  observe coarsening in the
limit $T  \to 0$, we  must let $t_p\rightarrow \infty$,  while keeping
$t_p\exp{(-4J/T)}$  fixed, and consider  domain diffusion  in rescaled
time $t =  t_p\exp{(-4J/T)}$. In this time variable,  the rate of spin
emission becomes of order unity, that of domain hops $\sim 1/l$, where
$l$  is a  typical domain  size,  and coarsening  proceeds via  domain
annihilation induced  by the diffusive  dynamics of the  domains.  The
complete  set of  microscopic spin  processes  has been  reduced to  a
size-dependent   domain  diffusion   model  in   time   $t$.   Certain
inconsistencies arise from this size-dependence for $l=1,2$, but these
`end-effects'  are  negligible  as far  as the  late-time  dynamics is
concerned  (as  is the  early-time  single-spin  regime, which  occurs
instantaneously  in terms of  the time  variable $t$).   Following the
introduction  of this  coarse-grained model,  we calculate  the domain
density  $N(t)$, and  conclude the  section with  a discussion  of the
persistence probability $Q(t)$.

\subsection{Domain density}
\label{domden}

\begin{figure}
\includegraphics[width=\linewidth]{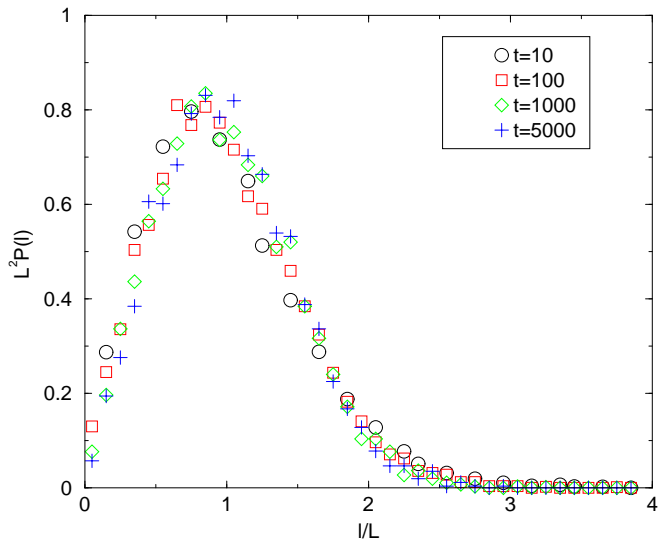}
\caption{Plot of  the scaling function $P_{sc}(x)$  against the scaled
length $x$ from numerical simulations.   Both the linear small-$x$ and
gaussian large-$x$ behaviour are reproduced.}
\label{sc}
\end{figure}

\begin{figure}
\includegraphics[width=\linewidth]{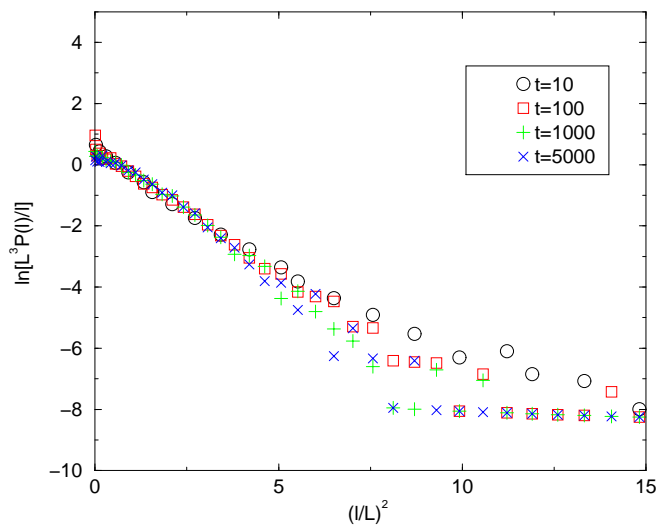}
\caption{The    data    of    Fig.\ref{sc}    plotted    instead    as
$\ln(P_{sc}x^{-1})$ versus $x^2$ }
\label{sc2}
\end{figure}

We  focus  first  on the  time  evolution  of  the domain  density  by
calculating $N(t)$, the total number  of domains per site at time $t$.
The  distribution  of domain  sizes,  $l$, at  time  $t$  is given  by
$P(l,t)$,  the number  of domains  of size  $l$ per  lattice  site. We
expect  that asymptotically  the  latter will  be  characterised by  a
single  dynamical length  scale, the  average domain  length $L(t)\sim
t^{\phi}$.  The scaling hypothesis for $P(l)$ is:
\begin{equation}
P(l,t)\sim t^{-2\phi} P_{sc}(x),\ \ \ x=\frac{l}{t^\phi}
\end{equation}
where   the  form  of   the  prefactor   is  fixed   by  magnetisation
conservation,   since  $\int_0^{\infty}   l  P(l)   dl$   must  remain
constant. The  $l$-dependence of  the distribution at  small $l$  is a
measure of  the availability  of domains near  the $l=0$  boundary and
dictates  the rate  of domain  extinction.  More  precisely,  a domain
disappears when it has  length $l=1$.  Therefore $dN/dt \propto P(1,t)
\to (dP/dl)|_{l=0}$ in  the continuum limit.  Since this  rate must be
finite, it follows  that $P(l,t) \propto l$ for $l \to  0$.  A plot of
$P_{sc}(x) =  L^2 P(l,t)$  against $x  = l/L$, where  $L$ is  the mean
domain   size,   confirms    the   linear   small-argument   behaviour
(Fig.\ref{sc})   whereas   the  same   data   plotted   in  the   form
$\ln(x^{-1}P_{sc})$  versus  $x^2$   (Fig.\ref{sc2})  shows  that  the
analytic form
\begin{equation}
P_{sc}(x) = A x \exp(-\lambda x^2)
\label{Gaussian} 
\end{equation}
provides a very good description of the data.

It  is straightforward  to generalize  the discussion  to the  case of
non-zero  magnetisation per spin  $\mu$. Then  the positive  ($+$) and
negative  ($-$)  domains will  have  different distribution  functions
$P_\pm(l_\pm,t)$.    Numerical   studies   confirm   that   the   form
(\ref{Gaussian})  provides a  good fit  to the  data for  $+$  and $-$
separately, but with different scale lengths $L_\pm(t)$. This suggests
the scaling forms
\begin{equation}
P_{\pm}(l_{\pm},t)=\alpha_{\pm}\frac{l_{\pm}}{L_{\pm}^{3}(t)}
F\left(\frac{l_{\pm}}{L_{\pm}(t)}\right)\ ,
\label{scaling}
\end{equation}
where  $F(x)$ is  defined such  that $F(0)=0$,  and  all normalisation
factors have been combined in the constants $\alpha_\pm$.

Returning to  the rate of decay of  N(t) we can write,  in the general
case,
\begin{eqnarray}
\frac{dN}{dt}&   =   &  -2\left(\left\langle   l_{-}^{-1}\right\rangle
\frac{\partial    P_{+}(l_{+})}{\partial   l_{+}}\mid_{l_{+}=0}\right.
\nonumber             \\             &&            +\left.\left\langle
l_{+}^{-1}\right\rangle\frac{\partial            P_{-}(l_{-})}{\partial
l_{-}}\mid_{l_{-}=0}\right)   \nonumber   \\   &=&-2\left(\left\langle
l_{-}^{-1}\right\rangle\frac{\alpha_{+}}{L_{+}^3}         +\left\langle
l_{+}^{-1}\right\rangle\frac{\alpha_{-}}{L_{-}^3}\right)\ .
\label{dndt}
\end{eqnarray}   
In the  above expression we have used  the fact that the  rate for the
elimination of  $+$ and $-$  domains with $l=1$  is given by  the mean
rate   at  which  $-$   and  $+$   domains  diffuse,   i.e.\  $\langle
l_{-}^{-1}\rangle$  and $\langle l_{+}^{-1}\rangle$  respectively. The
extra factor of 2 comes from the fact that the elimination of a domain
reduces the number of domains by 2 (since the domains neighbouring the
eliminated domains merge). For  an exact result, these averages should
be taken only over domains adjacent to domains of size $l=1$. In order
to make further progress, we relax this condition and take the average
over  all  domains  of  a  given size.  We  are  therefore  neglecting
correlations between domain sizes.

The required averages $\langle l_{\pm}^{-1}\rangle$ are given by
\begin{equation}
\langle                       l_{\pm}^{-1}\rangle=\frac{\int_0^{\infty}
l_{\pm}^{-1}P_{\pm}(l_{\pm})dl_{\pm}}{\int_0^{\infty}
P_{\pm}(l_{\pm})dl_{\pm}}={\Large \frac{2 \alpha_{\mp}}{N(t) L_{\pm}^2
}} f_0\ ,
\label{avll}
\end{equation}
where $f_n=\int_0^{\infty} x^n F(x) dx$ and $N(t)$ is the total number
of  domains per lattice  site. The  latter is  obtained by  adding the
contributions  $N_+(t)$   and  $N_-(t)$  from  $+$   and  $-$  domains
respectively:
\begin{eqnarray}
N(t)=\int_0^{\infty} P_{+}(l{+})dl_+ + \int_0^{\infty} P_{-}(l{-})dl_-
=\nonumber\\
=\left(\frac{\alpha_+}{L_+}+\frac{\alpha_-}{L_-}\right)f_{1}
=\frac{2\alpha_{+}f_{1}}{L_+},
\label{nn}
\end{eqnarray}
where the last equality follows from the condition $N_+(t)=N_-(t)$ for
all  $t$, and holds  for all  values of  the conserved  quantity $\mu$
defined as the fraction of spins  which have the value $+1$ (i.e.\ the
magnetisation per site is $2\mu-1$). This quantity satisfies
\begin{equation}
\mu=\int_0^{\infty} l_+ P_{+}(l_{+}) dl_+ = \alpha_+ f_2
\label{mu}
\end{equation}
and
\begin{equation}
1-\mu=\int_0^{\infty} l_- P_{-}(l_{-}) dl_- = \alpha_- f_2.
\label{mmu}
\end{equation}

Using Eqs.\ (\ref{avll}),(\ref{nn}), (\ref{mu}) and (\ref{mmu}) we can
express $\langle l_{\pm}^{-1}\rangle$, $L_{\pm}$ and $\alpha_{\pm}$ in
terms of  $N(t)$, $\mu$ and  the integrals $f_n$  ($n=0,1,2$). Putting
these  into Eq.\  (\ref{dndt})  and integrating  yields the  following
expression for the domain density :
\begin{equation}
N(t)=\left(\frac{3f_0f_2^{3}t}{8
f_1^{5}\mu^2(1-\mu)^2}+\frac{1}{N(0)^3}\right)^{-1/3}\ .
\end{equation}
It remains to  determine the constants $f_n$.  From  the definition of
the average domain lengths we find:
\begin{eqnarray}
L_{\pm}=\frac{\int_0^{\infty}
l_{\pm}P_{\pm}(l_{\pm})dl_{\pm}}{\int_0^{\infty}
P_{\pm}(l_{\pm})dl_{\pm}}=\frac{f_2}{f_1}L_{\pm}\nonumber\\
\Rightarrow f_2=f_1\ .
\label{f2f1} 
\end{eqnarray}
For the  assumed Gaussian scaling  function $F(x)=\exp(-\lambda x^2)$,
the    required     integrals    give    $f_0=(\pi/4\lambda)^{1/2}$,
$f_1=1/(2\lambda)$   and  $f_2=(\pi/16\lambda^3)^{1/2}$.   The  result
$f_2=f_1$ fixes $\lambda = \pi/4$, giving $f_0=1$ and $f_1=f_2=2/\pi$.
The final expression for N(t) becomes:
\begin{equation}
N(t)=\left(\frac{3\pi^{2}t}{32\mu^2(1-\mu)^2}
+\frac{1}{N(0)^3}\right)^{-1/3}\ .
\label{final}
\end{equation}

\begin{figure}
\includegraphics[width=\linewidth]{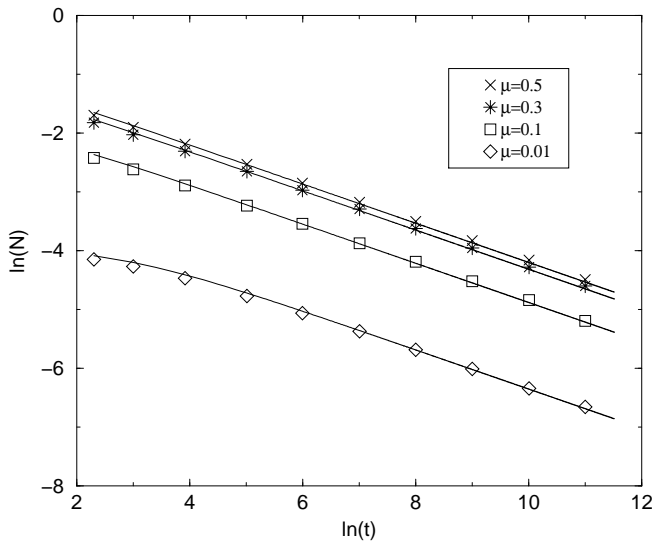}
\caption{Time-dependence of the total domain density, $N(t)$, for 
different volume fractions, $\mu$, of the $+$ phase. The continuous 
curves show the analytical prediction given by Eqn.\ (ref{final}).}
\label{figmu}
\end{figure}

In Figure  \ref{figmu} we have plotted  eq.(\ref{final}) for different
values  of  $\mu$  together   with  numerical  data  from  simulations
described in the following subsection.   The agreement is good even at
early times, verifying the fast collapse to scaling observed in Figure
\ref{sc}. Note  that in the  derivation of Eq.\ (\ref{final})  we used
the  scaling form  (\ref{scaling}) for  all times  $t$, not  just late
times. This may account for the small discrepancies between theory and
data evident at early times in Figure \ref{figmu}.  It is also evident
that the  data points lie  slightly above the  corresponding continuous
curves at  late times.  Nevertheless,  the generally good fit  of Eq.\
(\ref{final}) to  the data,  over a  wide range of  $\mu$ and  with no
adjustable  parameters, provides  good evidence  that  the assumptions
made  in its  derivation  are either  correct  or at  least very  good
approximations.  Principal among  these assumptions  are  the Gaussian
form  taken  for  the  scaling  function $F(x)$  (to  derive  explicit
expressions  for $f_0$,  $f_1$ and  $f_2$), the  assumption  that this
scaling function is the same for both domain types, and the neglect of
correlations between domain sizes.

To probe the quality of the fit more closely, we extract the amplitude
$C$ in  the asymptotic  behaviour $N(t) \to  Ct^{-1/3}$ for large $t$.
For $\mu=1/2$, we find $C=0.440  \pm 0.003$, whereas the analytic form
(\ref{final})  gives  $C=0.407$,  a  7.5\% discrepancy.  Ben-Naim  and
Krapivsky  \cite{krapivsky98} have given  an alternative  treatment of
the  domain-size distribution,  making  only the  assumption that  the
domain sizes are uncorrelated. This  leads to a slightly better value,
0.415, for  $C$. However,  no results were  given for other  values of
$\mu$ in  ref.\ \cite{krapivsky98}. Our data show  that the calculated
$C$ is smaller than the measured one by about 7.5\% for $\mu=0.3$ and
0.1, and by about 5\% for $\mu=0.01$.

\subsection{Persistence probability}
\label{persprob}

For the numerical simulations we adopted the domain description of the
spin  problem, since  it  is the  most  appropriate for  $T\rightarrow
0$. We worked  on 1D lattices of size  $n=10^5$ with periodic boundary
conditions. All  samples were prepared by randomly  assigning spins to
the  $n$ sites,  although  the system  can  be prepared  in any  other
homogeneous  initial  state.   The   early-time  form  of  the  domain
size-distribution is irrelevant to the asymptotic behaviour.

Updating  the  persistence   probability  $Q(t)$  means  tracking  the
fraction of spins that have not flipped or, equivalently, the fraction
of sites  that have not been  visited by a  domain wall up to  $t$. On
average, we expect each of the  $N$ domains to attempt to move by spin
emission once per unit time in agreement with the rescaled rates, so a
simple Monte-Carlo  algorithm would be  to select a domain  at random,
move it with  probability $1/l$ in either direction,  where $l$ is the
its length,  and increment time  by $\Delta t=1/N(t)$  irrespective of
whether or not the domain is moved.  This is indeed the method used in
\cite{sire98} to infer the value of the persistence exponent, 
$\theta=0.73$. 

Unfortunately this  algorithm suffers from severe  slowing-down in the
scaling limit, where the average domain length $L$ is much larger than
unity and  most hopping  attempts are unsuccessful.  We worked  with a
modified version of  the domain algorithm which does  not exhibit such
problems by  ensuring that  a move is  accepted at  each computational
step.  A normalized  vector is constructed for the  $N$ domains, whose
$i^{th}$  element, $H_i(t)=l_i^{-1}/(\sum_{i=1}^N  l_i^{-1})$,  is the
probability that  domain $i$ experiences a  hop at time  $t$.  For any
{\em given} hop this probability is size-dependent only.  Incrementing
time  by $1/  \sum_{i=1}^N l_i^{-1}$,  the average  time  between {\em
successful} moves,  we end up with  the same diffusion rate  as in the
simpler  algorithm.  We  avoid,  however,  the  problematic  behaviour
relating to the inefficiency of the $1/l$ trial. The disadvantage here
is the calculation of the vector  ${\bf H}$ at early times, which does
result in the simpler algorithm  being faster in this regime, but this
is a  small price to  pay. We show  in section \ref{gamma2}  that slow
convergence to a scaling state  in systems with diffusion rates $D\sim
l^{\gamma}$  with   $\gamma<-1$  reinforces  the  need   for  the  new
algorithm.

\begin{figure}
\includegraphics[width=\linewidth]{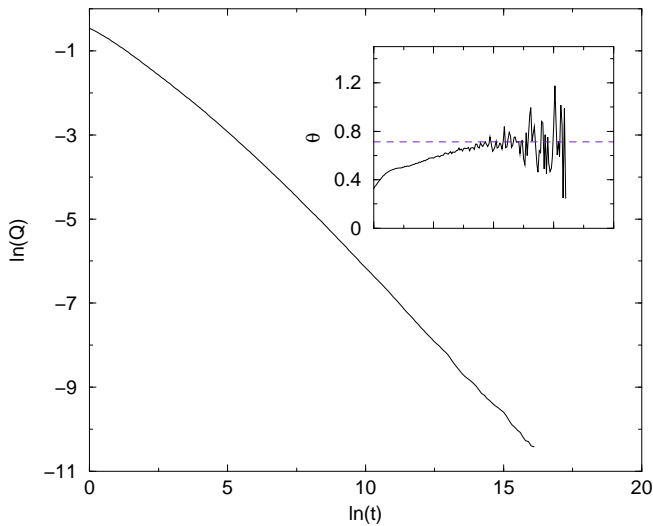}
\caption{Logarithm    of   the    persistence    probability   against
$t' = \ln t$. The  local slope  $\theta(t')$ is also  given in
the inset.}
\label{g-1}
\end{figure}

In  Figure \ref{g-1}  we present  a  log-log plot  of the  persistence
probability against time, averaged over 94 realisations. The effective
exponent $\theta(t)=-d\ln  Q/d\ln t$, plotted  against $\ln t$  in the
inset of  Fig.\ ref{g-1},  seems to saturate  at value  $\theta \simeq
0.716$,  the spread  around  this value  beyond  $t\sim e^{11}$  being
primarily due  to poor statistics  resulting from the small  number of
persistent sites remaining at this timescale.

We  note  that,  since  $\theta   \approx  0.7$  is  larger  than  the
domain-growth  exponent  ($\phi=1/3$),  the typical  distance  between
persistent  sites, $t^\theta$,  grows faster  than the  domain length,
$t^\phi$.  Clusters   of  non-persistent  sites,   of  typical  length
$t^\theta$,  grow  by  the  motion   of  the  domain  walls  at  their
boundaries.  These  motions  are  uncorrelated since  the  separation,
$t^\theta$,  of the  relevant walls  is  much larger  than the  domain
scale.   The   statistical   independence  of   persistence-destroying
processes suggests further that  the distribution function, $n(s)$, of
the size $s$ of  non-persistence intervals should decay exponentially,
$n(s) \propto  \exp(-s/\langle s  \rangle)$, where $\langle  s \rangle
\sim  t^\theta$  is  the   mean  size  of  a  non-persistent  interval
\cite{manoj00}. This  is to be  contrasted with the Glauber  model, in
which the  persistence length, $t^{3/8}$,  is smaller than  the domain
scale, $t^{1/2}$, and $n(s)$ has a nontrivial form \cite{manoj00}.

\section{Conserved dynamics in aggregation}
\label{aggg}

In this  section we  compare the coarsening  dynamics of  the Kawasaki
model  to  that  of  aggregation  models.  How  would  the  coarsening
properties be modified in the  Kawasaki ferromagnet for choices of the
diffusion exponent  $\gamma$ other  than $\gamma =  -1$ in  $D\sim D_0
l^{\gamma}$?      Furthermore,     is     the      amplitude     ratio
$R_0=D_{0}^{(+)}/D_{0}^{(-)}$  of the diffusion  constants of  the $+$
and $-$  domains relevant to  the persistence probability and,  if so,
will a  distinction between `up' and `down'  persistence be necessary?
We  discuss these  points with  the aid  of results  from irreversible
cluster-cluster  aggregation and  extensive  numerical simulations  of
modified versions of the Kawasaki system. For the rest of the paper we
shall use  `Kawasaki' to denote a  more flexible model  where we allow
$\gamma\ne-1$ and $R_0\ne 1$. It must be pointed out, that both $D\sim
1/l$ and particularly $D_{0}^{(+)}=D_{0}^{(-)}$ are robust features of
Kawasaki  dynamics, which  follow  directly from  the elementary  spin
processes.  This  follows most strongly  from the symmetric  nature of
the emission process where both an  `up' and a `down' spin are emitted
during  a  single  event  and  consequently  $R_0=1$  is  unavoidable.
Although   it  is  not   possible  to   reconcile  these   rules  with
spin-exchange  at  the microscopic  level,  changes  in the  diffusion
exponent and  the ratio of  amplitudes are easily accommodated  in the
domain description that we have employed so far.

\subsection{Diffusion-limited cluster-cluster aggregation (DLCA)}

Irreversible  cluster  aggregation  is  closely  related  to  Kawasaki
dynamics.  It, too,  involves  the separation  of a  two-phase mixture
through domain  diffusion and leads to conservation  of its respective
order parameter, in this case the total mass $m$. Clusters, defined as
intervals of  consecutive occupied  sites, move randomly  and coalesce
irreversibly  on impact,  whilst the  `empty' phase  through  which 
the clusters  diffuse  is   immobile (reversible aggregation scenarios,
including  various  fragmentation  mechanisms  or  monomer  `chipping'
\cite{majumdar},  can actually  result in  effective diffusion  of the
empty  phase,  and hence  partly  restore  the dynamic  equivalence).
Consequently,  this  model   corresponds  to  the  $R_0=0$  `Kawasaki'
system. The  diffusion rate for a  cluster of length $l$  is $D_F \sim
l^{\gamma}$  but   we  shall  concentrate   on  $\gamma  \le   0$  and
particularly  on  $\gamma=0,-1$, which  correspond  to artificial  but
instructive realizations of domain diffusion on the Ising lattice.

Due to the broken mobility symmetry between the `filled' (cluster) and
`empty' phases,  a distinction should be made  between the persistence
properties   of   initially   occupied   sites  and   initially  empty
ones.  Therefore,   we  use  $P_E(t)$  to   denote  the  persistence
probability of empty  sites and $P_F(t)$ that  of occupied sites with
similar notation  applying to the corresponding  exponents. Only $P_E$
is  universal whereas $P_F$  depends strongly  on the  volume fraction
$\mu$  of  the  `filled'  phase,  which  of  course  remains  constant
throughout the coarsening process.

\begin{figure}
\includegraphics[width=\linewidth]{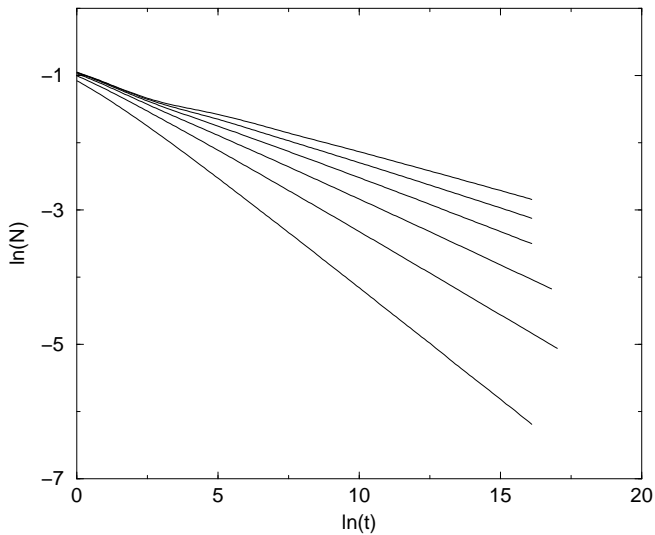}
\caption{Log-log plot of the domain density with time  for 
$\gamma=-6$ to $-1$ (top to bottom). The asymptotic  slopes are
-0.117,-0.136,-0.162,-0.197,-0.248 and -0.332 respectively, in 
reasonable agreement with the prediction $\phi = 1/(2-\gamma)$.}
\label{wg}
\end{figure}

For negative  $\gamma$ the exponents  governing the algebraic
decay of  the cluster density  and empty persistence  probability have
been  calculated \cite{hellen01}  to  be $\phi^{cc}=1/(2-\gamma)$  and
$\theta_{E}=2 \phi^{cc}$  for the DLCA  model.  The dependence  of the
dynamic exponent on $\gamma$  is readily obtainable from scaling alone
\cite{hellen00},  an  approach whose  validity  extends to  `Kawasaki'
dynamics in view of sec.\ref{domden}.  If the domain scale is $l$, the
time  required  to  develop  this  scale  is  $t  \sim  l^2/D(l)  \sim
l^{2-\gamma}$, whence $l  \sim t^{1/(2-\gamma)}$. For the empty-site
persistence, i.e.\  the probability that  an initially empty  site has
not  been covered by  a cluster  up to  time $t$,  Helen et  al.\ used
$D(t)=D_0  t^{\gamma\phi}$  as  a  mean-field  representation  of  the
cluster kinetics to  get $\theta_{E}$, which is known  to be exact for
$\gamma=0$.  Numerical simulations indicate   that the  persistence of
the  `filled' phase  also  decays algebraically  with a  non-universal
exponent  $\theta_{F}(\mu)$  and  that  $\theta_{F}=\theta_E$  in  the 
half-filled case, $\mu=1/2$, for all $\gamma$.
 
\begin{figure}
\includegraphics[width=\linewidth]{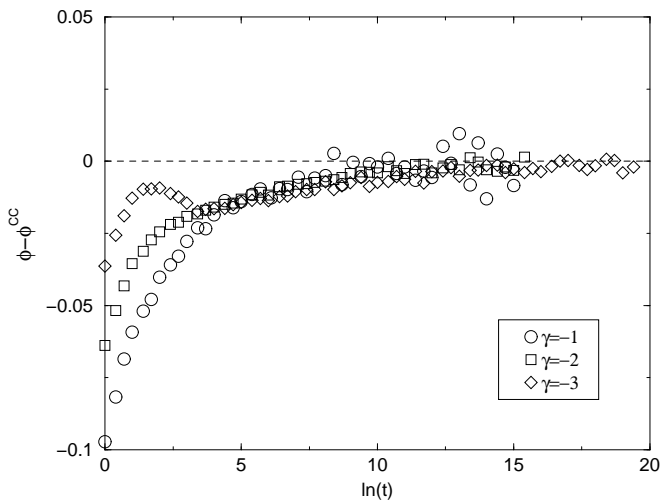}
\caption{Plot      of     the      difference     $\phi-\phi^{cc}=\phi
(R_0=1)-\phi(R_0=0)$ in logarithmic time, for 3 values of $\gamma$.}
\label{twg}
\end{figure}

\begin{figure}
\includegraphics[width=\linewidth]{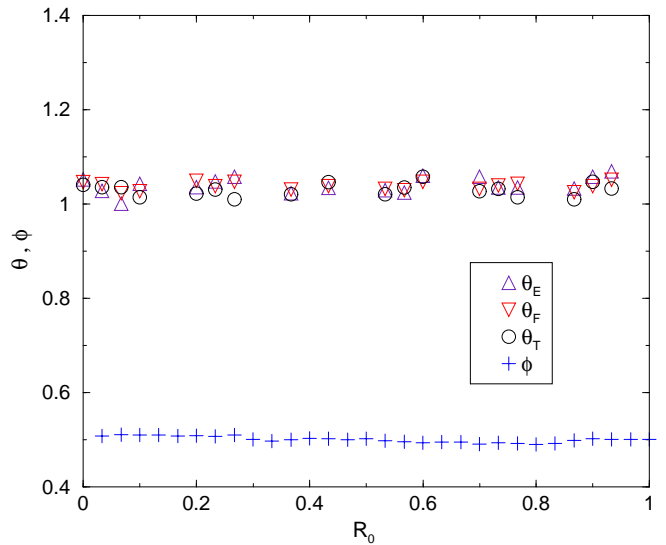}
\caption{Dynamic  and persistence  exponents for  the size-independent
`Kawasaki'  model  in the  range  $0\le  R_0\le  1$.  We  have  chosen
$D_E=D_+$. The upper and lower data sets give the persistence and growth 
exponents, $\theta$ and $\phi$, respectively; $\theta_T$  refers to the 
persistence of  the whole lattice and is obtained from the sum of the 
two constituent probabilities. The values of the exponents are consistent 
with $\phi^{cc}=1/2$ and $\theta^{cc}=1$. }
\label{thph0}
\end{figure}

Agreement with simulations  of the zero-magnetisation `Kawasaki' model
is  very good  for the  form  of the  dynamic exponent  in the  regime
$-1\le\gamma\le -6$  (Fig.\ref{wg}).  The  time required to  reach the
asymptotic value seems independent  of the diffusion exponent $\gamma$
(Fig.\ref{twg}),  so  that all  $\theta(\gamma)$  saturate within  the
numerical window.   The data suggest  that domain growth  is identical
for   the   limiting   cases   $R_0=0$   (aggregation)   and   $R_0=1$
('Kawasaki'). We believe this to be  the case in the whole range $0\le
R_0  \le  1$,  given  eq.(\ref{dndt}) and  simulations  indicating  no
systematic  change  in   $\phi(\gamma=0)$  for  constant  but  unequal
diffusion coefficients (Fig.\ref{thph0}).

\subsection{The cases $\gamma=0,-1$}
\label{gamma1}

\begin{figure}
\includegraphics[width=\linewidth]{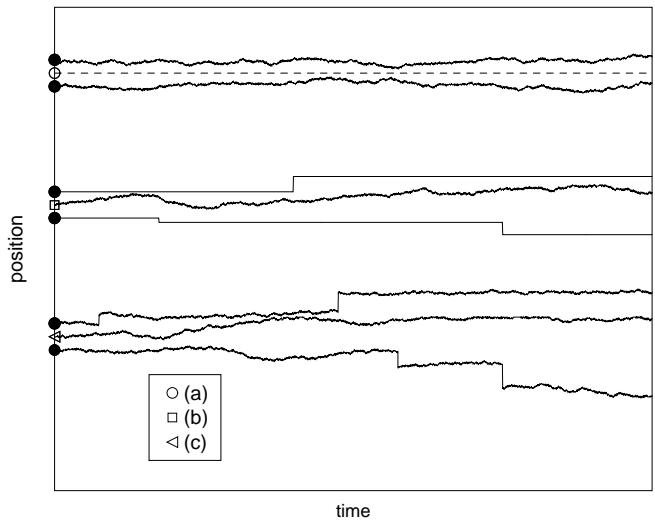}
\caption{Space-time diagram of persistence-destroying processes in 
(bottom to top) (a) empty, (b) filled and (c) symmetric phases.}
\label{prpr}
\end{figure}

Let  us  examine size-independent  diffusion  ($\gamma=0$) first.  The
empty-site persistence probability for $R_0=0$ reduces to the survival
probability of a stationary point  surrounded by two random walkers --
the   near  edges   of   the  two   domains   surrounding  the   point
(Fig.\ref{prpr}(a)).   This  survival  probability  is  given  by  the
product of the  survival probability due to each  walker and therefore
decays  asymptotically as $t^{-1/2}t^{-1/2}  = t^{-1}$.  Similarly, we
can interpret filled-site persistence as the survival probability of a
random  walker  with   receding  boundaries  (Fig.\ref{prpr}(b)).  The
persistent site carries the diffusion  of the domain in which it lies,
while the edges expand away from  it due to domain collisions. For any
other $R_0$ the survival of  the persistent site involves a mixture of
both processes (Fig.\ref{prpr}(c)).

Although the  exact kinetics of individual  walls in Fig.\ref{prpr}(c)
remain  unknown,  hints  as  to   their  form  have  emerged  in  many
contexts.  In a  study of  annihilating random  walks on  the infinite
half-line \cite{krapivsky97} it  was shown from numerical measurements
that   the  probability   density   of  the   outermost  particle   is
asymmetric. In that case too the trajectory of the outermost walker is
a combination of diffusion  and directed jumps resulting from particle
annihilation,  similarly to  Kawasaki walls.  The  probability density
curve was found  to decay like $\exp{(-\alpha x^2)}$  in the direction
of the free interface, but like  $\exp{(-\beta x)}$ on the side of the
filled lattice, a behaviour which  seems to suggest a regime where the
diffusion  exponential  dominates   and  one  where  the  annihilation
exponential  does.  This appears  reasonable  on  the  basis that  the
contribution of these irregular hops is super-diffusive, characterised
by the domain  length distribution from which the  jumps are drawn. In
studies of the $A+A\rightarrow  0$ reaction, this component has indeed
been shown  to exhibit an  exponential form of the  kind $\exp{(-\beta
x)}$\cite{derrida96}.  Unfortunately,  our   calculation  on  a  naive
convolution of  the two components  has turned out an  overestimate of
the  persistence exponent  in the  case where  the persistent  site is
stationary and the probability can  be expressed as the product of the
two  individual capture  probabilities,  suggesting that  correlations
between the diffusion and annihilation processes cannot be neglected.

As  far as numerical  simulations of  the size-independent  system are
concerned, we have  found that within the accuracy  of the statistics,
no systematic shift in the values  of the exponents takes place in the
transition  from irreversible aggregation  to symmetric  diffusion. It
should  be noted that  the data  in Fig.\ref{thph0}  consistently fall
slightly above  the predicted value $\theta=1$. Since  the exponent is
known   to  be   1  in   the  limiting   cases  $R_0=1$   and  $R_0=0$
\cite{hellen01} it may  be that the numerics have  not yet reached the
true  asymptotic  value.  Nevertheless,  even considering  this  small
discrepancy  along all values  of $R_0$,  $\gamma=0$ remains  the only
case for  which the values of  all four exponents match  in the limits
$R_0=0$  and $R_0=1$.  For Kawasaki  dynamics ($\gamma=-1$),  which we
discussed  in   Sec.\ref{persprob},  a  deviation   of  $\theta$  from
$\theta_{cc}$ is visible and begins to grow as we reduce the diffusion
exponent  further   (see,  Sec.\ref{gamma2}).  The   inset  in  Figure
\ref{g-1} indicates  that $\theta$ saturates  asymptotically above the
$2/3$ aggregation value, our calculations favouring $\theta=0.716(5)$.

\subsection{$-2\ge\gamma\ge -6$}
\label{gamma2}

\begin{figure}
\includegraphics[width=\linewidth]{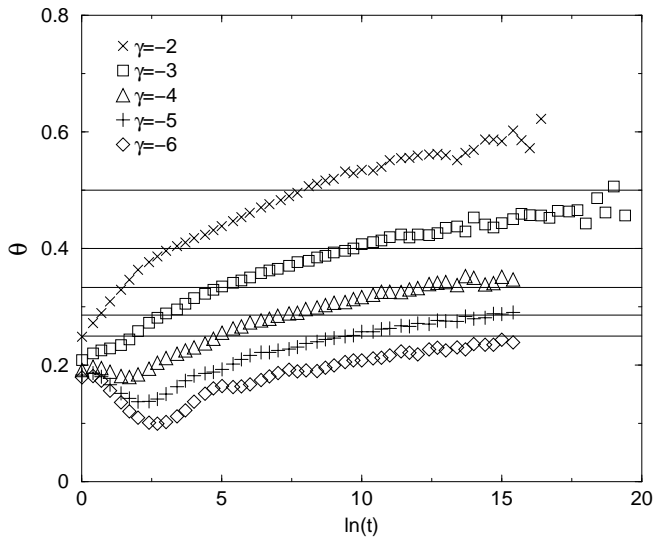}
\caption{Effective persistence  exponent against  $\ln t$  for  
$R_0=1$  `Kawasaki' systems with negative $\gamma$.   The  horizontal 
lines  mark the  values $\theta^{cc}$ predicted for $R_0=0$.}
\label{tpall}
\end{figure}

In this  parameter range, the  aggregation and `Kawasaki'  results for
the persistence  exponent clearly  disagree.  Our simulations  did not
reach,  in  most   cases,  the  asymptotic  regime  (Fig.\ref{tpall}).
Nevertheless,  they demonstrate that  the exponents  invariably exceed
the aggregation  prediction $\theta_{cc}=2/(2-\gamma)$. The  very slow
convergence we observe  has, in fact, also arisen  in other studies of
aggregation-fragmentation models \cite{rajesh02}, a class to which the
`Kawasaki'  model can  be shown  to relate.   It was  noted  there, in
numerical estimates  of the  cluster mass distribution  function, that
for $\gamma<-2$ simulations failed to  approach a scaling state in the
time  available.   Our own  efforts  led us  in  many  cases to  final
configurations  where the average  domain size  was comparable  to the
lattice  size.   We  think  that  for  the  algorithm  employed,  good
scaling-state statistics would require  unfeasibly long runs. We found
no  obvious way in  which to  improve further  the performance  of the
routine which is still limited by the update of the probability vector
$H$,   although   this  penalty   becomes   less   relevant  at   late
times. Unfortunately, lattice  renormalisation techniques (such as the
one  in  \cite{ben-avraham}) which  have  proven  successful in  other
circumstances  can not  be used  in this  case as  they lead  to local
violations of the conservation principle.

It is also clear that the  collapse onto a single curve encountered in
Figure  \ref{twg}   at  large  $t$  is  absent   for  the  persistence
exponent.  This  much  slower  approach to  asymptotic  behaviour  for
$\theta$ than for $\phi$ is also consistent with the statement made in
Sec.\ref{persprob}  that non-persistent  intervals and  domain lengths
have different scaling characteristics.

Despite these complications, lower limits for $\theta(\gamma)$ are set
by the data, and imply  faster-than-$\theta^{cc}$ decay for $Q$ in the
symmetric system.  This  is perhaps surprising in view  of the results
in the  previous subsection, where  such a deviation from  the $R_0=0$
behaviour is absent. We do not know of any argument for the $\gamma=0$
system to constitute a special  case, but it is tempting to conjecture
that  the deviation of  $\theta$ from  $\theta^{cc}$ observed  for all
$\gamma<-1$ is monotonic in the intermediate regime $0<R_0<1$.

\section{Discussion and Summary}

In  this  paper we  considered  an  integrated  analysis of  two-phase
conserved systems in one dimension based on two examples: irreversible
cluster-cluster aggregation,  where only one of the  phases is allowed
to diffuse,  and Kawasaki dynamics,  where both phases diffuse  at the
same rate.  In  terms of the ratio of  diffusion amplitudes $R_0$, the
algebraic  domain  growth  at  late  times  has  been  established  as
identical  in  the  these  two  limits, with  the  `Kawasaki'  dynamic
exponent  $\phi$ matching the  prediction $1/(2-\gamma)$  whatever the
form of size-dependence assumed for the diffusion rate. In the case of
size-independent diffusion,  we have also found  that this equivalence
holds  for all  other intermediate  values of  $R_0$, as  was expected
given  the  restrictions  on   scaling  imposed  by  the  conservation
principle,  i.e.\  that the  order  parameter  should remain  constant
throughout  coarsening.  The robustness  of  this  behaviour has  been
confirmed     in     a      variety     of     aggregation     systems
\cite{rajesh02,hellen01}.

By contrast,  in the critical  $\mu=1/2$ limit considered,  changes in
the relative mobility of the  two phases lead to different persistence
characteristics.   The  transition  from irreversible  aggregation  to
symmetric  diffusion  is accompanied  by  severe  slowing-down in  the
convergence  rate  of  the   effective  persistence  exponent  to  its
asymptotic  value  as  well  as  a  distinct  shift  from  the  simple
$\theta^{cc}=2/(2-\gamma)$ result  for all negative  $\gamma$. We have
been able  to show  that the disagreement,  which is at  zero $\gamma$
absent  in the whole  range $0\le  R_0\le 1$,  always leads  to faster
persistence  decay  for  the  symmetric system,  with  numerical  data
indicating a possible lower bound for the exponent at $\theta=2.3\phi$
$ (\gamma<-2)$.

Apart from the slow  convergence to asymptotic behaviour which hinders
any  definitive results  as far  as  the  value  of $\theta$  in  the
symmetric  limit   is  concerned,  it  is  also   possible  that,  for
size-dependent  cluster dynamics  and finite  values of  $R_0$, $Q(t)$
deviates from the simple  power-law behaviour observed in irreversible
aggregation.   In   this  latter  case,  it  seems   that  the  filled
persistence (the persistence of the mobile phase) is non-universal and
decays  as  a power-law  only  for  $\mu=1/2$ \cite{hellen01}.   Since
single-site  persistence for $R_0>0$  involves a  mixture of  both the
empty- and  filled-site behaviour, it  is possible that  the resulting
probability is also non-universal. 

A  further prospect  which must  be considered  is the  possibility of
inferring $\theta$  in the  more strongly size-dependent  systems from
the  scaling  properties of  the  size-distribution of  non-persistent
sites. The general characteristics of this scaling are dictated by the
fact   that  $\theta>\phi$,   which   constrains  the   tail  of   the
non-persistent   interval   size-distribution    to   a   known   form
\cite{manoj00}.  Hints  of this behaviour  have been suggested  by our
results for the probability $Q(t)$, since the regularity encountered in
the collapse  of the growth  exponent onto the aggregation  result for
different $\gamma$ is lost in  plots of $\theta$. Instead, a very slow
saturation  of  the  local  persistence  exponent  has  been  observed
extending,  in all cases  but $\gamma=-1$, well  beyond $t\sim e^{15}$.

\section{ACKNOWLEDGEMENT}

The work of PG was supported by EPSRC.


\begin{thebibliography}{30}
\expandafter\ifx\csname natexlab\endcsname\relax\def\natexlab#1{#1}\fi
\expandafter\ifx\csname bibnamefont\endcsname\relax
  \def\bibnamefont#1{#1}\fi
\expandafter\ifx\csname bibfnamefont\endcsname\relax
  \def\bibfnamefont#1{#1}\fi
\expandafter\ifx\csname citenamefont\endcsname\relax
  \def\citenamefont#1{#1}\fi
\expandafter\ifx\csname url\endcsname\relax
  \def\url#1{\texttt{#1}}\fi
\expandafter\ifx\csname urlprefix\endcsname\relax\def\urlprefix{URL }\fi
\providecommand{\bibinfo}[2]{#2}
\providecommand{\eprint}[2][]{\url{#2}}


\bibitem{bray94}
\bibinfo{author}{\bibfnamefont{A.~J.} \bibnamefont{Bray}},
\bibinfo{journal}{Adv.\ Phys.} \textbf{\bibinfo{volume}{43}},
\bibinfo{pages}{357} (\bibinfo{year}{1994}).

\bibitem{derrida94}
\bibinfo{author}{\bibfnamefont{B.}~\bibnamefont{Derrida}},
\bibinfo{author}{\bibfnamefont{A.~J.} \bibnamefont{Bray}},
\bibinfo{author}{\bibfnamefont{C.}~\bibnamefont{Godreche}},
\bibinfo{journal}{J.\ Phys.\ A} \textbf{\bibinfo{volume}{27}},
\bibinfo{pages}{357} (\bibinfo{year}{1994}).

\bibitem{derrida95}
\bibinfo{author}{\bibfnamefont{B.}~\bibnamefont{Derrida}},
\bibinfo{author}{\bibfnamefont{V.}~\bibnamefont{Hakim}},
\bibinfo{author}{\bibfnamefont{V.}~\bibnamefont{Pasquier}},
\bibinfo{journal}{Phys.\ Rev.\ Lett.} \textbf{\bibinfo{volume}{75}},
\bibinfo{pages}{751} (\bibinfo{year}{1995}).

\bibitem{majumdar}
\bibinfo{author}{\bibfnamefont{S.~N.}~\bibnamefont{Majumdar}},
Curr.\ Sci.\ India {\bf 77}, 370 (1999), and cond-mat/9907407.  

\bibitem{cornell96}
\bibinfo{author}{\bibfnamefont{S.~J.}~\bibnamefont{Cornell}},
\bibinfo{author}{\bibfnamefont{A.~J.} \bibnamefont{Bray}},
\bibinfo{journal}{Phys.\ Rev.\ E} \textbf{\bibinfo{volume}{54}},
\bibinfo{pages}{1153} (\bibinfo{year}{1996}).

\bibitem{avraham}
\bibinfo{author}{\bibfnamefont{D.}~\bibnamefont{ben-Avraham}},
\bibinfo{author}{\bibfnamefont{F.}~\bibnamefont{Leyvraz}},
\bibinfo{author}{\bibfnamefont{S.}~\bibnamefont{Redner}},
\bibinfo{journal}{Phys.\ Rev.\ E} \textbf{\bibinfo{volume}{50}},
\bibinfo{pages}{1843} (\bibinfo{year}{1994}).

\bibitem{sire98}
\bibinfo{author}{\bibfnamefont{S.}~\bibnamefont{Cueille}},
\bibinfo{author}{\bibfnamefont{C.}~\bibnamefont{Sire}},
\bibinfo{journal}{Eur.\ Phys.\ J.\ B} \textbf{\bibinfo{volume}{7}},
\bibinfo{pages}{111} (\bibinfo{year}{1999}).

\bibitem{hellen01}
\bibinfo{author}{\bibfnamefont{E.~K.~O.} \bibnamefont{Hellen}},
\bibinfo{author}{\bibfnamefont{M.~J.} \bibnamefont{Alava}},
\bibinfo{journal}{Phys.\ Rev.\ E} \textbf{\bibinfo{volume}{66}},
\bibinfo{pages}{26120} (\bibinfo{year}{2001}).

\bibitem{cornell91}
\bibinfo{author}{\bibfnamefont{S.~J.} \bibnamefont{Cornell}},
\bibinfo{author}{\bibfnamefont{K.}~\bibnamefont{Kaski}},
\bibinfo{author}{\bibfnamefont{R.~B.} \bibnamefont{Stinchcombe}},
\bibinfo{journal}{Phys.\ Rev.\ B} \textbf{\bibinfo{volume}{44}},
\bibinfo{pages}{12263} (\bibinfo{year}{1991}).

\bibitem{krapivsky98}
\bibinfo{author}{\bibfnamefont{E.}~\bibnamefont{Ben-Naim}},
\bibinfo{author}{\bibfnamefont{P.~L.} \bibnamefont{Krapivsky}},
\bibinfo{journal}{J.\ Stat.\ Phys.} \textbf{\bibinfo{volume}{93}},
\bibinfo{pages}{583} (\bibinfo{year}{1998}).

\bibitem{jepessen}
\bibinfo{author}{\bibfnamefont{C.}~\bibnamefont{Jepessen}},
\bibinfo{author}{\bibfnamefont{O.~G.} \bibnamefont{Mouritsen}},
\bibinfo{journal}{Phys.\ Rev.\ B} \textbf{\bibinfo{volume}{47}},
\bibinfo{pages}{14724} (\bibinfo{year}{1993}).

\bibitem{hellen00}
\bibinfo{author}{\bibfnamefont{E.~K.~O.} \bibnamefont{Hellen}},
\bibinfo{author}{\bibfnamefont{T.~P.} \bibnamefont{Simula}},
\bibinfo{author}{\bibfnamefont{M.~J.} \bibnamefont{Alava}},
\bibinfo{journal}{Phys.\ Rev.\ E} \textbf{\bibinfo{volume}{62}},
\bibinfo{pages}{4752} (\bibinfo{year}{2000}).

\bibitem{rajesh02}
\bibinfo{author}{\bibfnamefont{R.}~\bibnamefont{Rajesh}},
\bibinfo{author}{\bibfnamefont{D.}~\bibnamefont{Das}},
\bibinfo{author}{\bibfnamefont{B.}~\bibnamefont{Chakraborty}},
\bibinfo{author}{\bibfnamefont{M.}~\bibnamefont{Barma}},
\bibinfo{journal}{Phys.\ Rev.\ E} \textbf{\bibinfo{volume}{66}},
\bibinfo{pages}{56104} (\bibinfo{year}{2002}).

\bibitem{godreche02}
\bibinfo{author}{\bibfnamefont{G.}~\bibnamefont{DeSmedt}},
\bibinfo{author}{\bibfnamefont{C.}~\bibnamefont{Godreche}},
\bibinfo{author}{\bibfnamefont{J.~M.} \bibnamefont{Luck}},
\bibinfo{journal}{Eur.\ Phys.\ J.\ B} \textbf{\bibinfo{volume}{27}},
\bibinfo{pages}{363} (\bibinfo{year}{2002}).

\bibitem{manoj00}
\bibinfo{author}{\bibfnamefont{G.}~\bibnamefont{Manoj}},
\bibinfo{author}{\bibfnamefont{P.}~\bibnamefont{Ray}},
J.\ Phys.\ A {\bf 33}, 5489 (2000); S. J. O'Donoghue and A. J. Bray, 
Phys.\ Rev. E, {\bf 62}, 3366 (2000).

\bibitem{krapivsky97}
\bibinfo{author}{\bibfnamefont{L.}~\bibnamefont{Frachebourg}},
\bibinfo{author}{\bibfnamefont{P.~L.} \bibnamefont{Krapivsky}},
\bibinfo{author}{\bibfnamefont{S.}~\bibnamefont{Redner}},
\bibinfo{journal}{J.\ Phys.\ A} \textbf{\bibinfo{volume}{31}},
\bibinfo{pages}{2791} (\bibinfo{year}{1998}).

\bibitem{derrida96}
\bibinfo{author}{\bibfnamefont{B.}~\bibnamefont{Derrida}},
\bibinfo{author}{\bibfnamefont{R.}~\bibnamefont{Zeitak}},
Phys.\ Rev.\ E {\bf 54}, 2513 (1996).

\bibitem{ben-avraham}
\bibinfo{author}{\bibfnamefont{D.}~\bibnamefont{ben-Avraham}},
\bibinfo{journal}{J.\ Chem.\ Phys.} \textbf{\bibinfo{volume}{88}},
\bibinfo{pages}{941} (\bibinfo{year}{1988}).


\end{thebibliography}
\end{document}